\documentclass[acmsmall,fleqn,screen]{acmart}
\acmConference[CHI'21 IICW]{ ACM CHI Conference on Human Factors in Computing Systems}{May 2021}{Online Virtual Conference}

\usepackage{subfigure}

\AtBeginDocument{%
  \providecommand\BibTeX{{%
    \normalfont B\kern-0.5em{\scshape i\kern-0.25em b}\kern-0.8em\TeX}}}

\begin{document}

\title[]{Insect-Computer Hybrid Speaker: Speaker using Chirp of the Cicada Controlled by Electrical Muscle Stimulation}

\author{Yuga Tsukuda}
\affiliation{%
  \city{Tsukuba}
  \institution{Research and Development Center for Digital Nature, University of Tsukuba}\country{Japan}}

\author{Naoto Nishida}
\affiliation{%
  \city{Tsukuba}
  \institution{University of Tsukuba}\country{Japan}}
  
\author{Jun-Li Lu}
\affiliation{%
  \city{Tsukuba}
  \institution{Research and Development Center for Digital Nature, University of Tsukuba} \country{Japan}}

\author{Yoichi Ochiai}
\affiliation{%
  \city{Tsukuba}
  \institution{Research and Development Center for Digital Nature, University of Tsukuba} \country{Japan}}
%\email{wizard@slis.tsukuba.ac.jp}

\renewcommand{\shortauthors}{Yuga Tsukuda et al.}

\begin{abstract}
We propose ``Insect--Computer Hybrid Speaker'', which enables us to make music made from combinations of computer and insects.
Many studies have proposed methods and interfaces to control insects and obtain feedback.
However, there has been less research on the use of insects for interaction with third parties.
In this paper, we propose a method in which cicadas are used as speakers triggered by using Electrical Muscle Stimulation (EMS).
We explored and investigated the suitable waveform of the chirp to be controlled, the appropriate voltage range, and the maximum pitch at which cicadas can chirp.
\end{abstract}

% 我々は，コンピュータと昆虫を組み合わせて音楽を再生させることが可能なスピーカ「Insect-Computer Hybrid Speaker」を提案する.

%We propose "Insect-Computer Hybrid Speaker", which enables us to make musics from combinations of computer and insects.

% 昆虫を制御する手法の研究、Userが昆虫を制御するインターフェースの研究、制御に加え昆虫からUserへのフィードバックを可能にする研究がされてきた．

% Lots of studies have proposed methods and interfaces for controlling insects and obtaining feedback.

% しかしながら，昆虫を通じた第三者へのインタンラクションに用いる試みはされてきていない．
% However, there have been no research on the usage of insects for interactions with third parties.
% However, previous researches were only focused between users and the insects. There has been no research, in which insects are used as devices on purpose of interactions with third parties.

% この論文では，Electrical Muscle Stimulation (EMS)によって，発音筋を制御し，蝉をスピーカとして使用する手法を提案する.
%In this paper, We propose a method in which cicadas are used as speakers with Electrical Muscle Stimulation(EMS).

% 制御に最適な鳴き声の種類を決定し、適切な電圧値、セミの出すことのできる音域を明らかにした。
% We explored and determined the best waveform of chirp to control, the appropriate voltage range, and the maximum pitch at which cicadas can chirp.



\maketitle

\section{Introduction}

There is an idea that integrates intersects and computers, namely 'Biobot'.  Biobots have attracted interest in the field of rescue and exploration \cite{DBLP:journals/computer/BozkurtLS16} due to their flexible mobility, durability, and supreme efficiency in energy resources. 
However, in the literature \cite{DBLP:journals/computer/BozkurtLS16, DBLP:journals/bc/RomanoDBS19, DBLP:conf/ssrr/WhitmanZTC18, doi:10.1098/rsif.2016.0060, 10.1371/journal.pone.0151808}, little research has been conducted on interactions between people using the insect-computer hybrid interface. In these investigations, an approach for interaction has utilized sounds. 
By attaching a speaker to the insects, the user can interact with other people. 
However, this method makes the inorganic part of the interface heavier and, as a result, hinders the movement of insects.
The solution to this problem is to use the chirps of the insects themselves.
These studies in anatomy and neurology have been conducted to control insect chirps \cite{10.2307/24950141, Bennet-Clark1681}.
However, only the simple chirping of an insect itself may not be enough to conduct complex interactions for people with quality. 
Especially, re-producing complex human voices by using the chirping of insects is difficult.
Therefore, we propose an insect-computer hybrid speaker that controls the pitch of insect chirps using an EMS.
In particular, we investigate whether the music can be composed by using insects.
We utilize cicadas as the devices to produce sounds since cicadas are easily controlled due to the fact that there are no large muscles or internal organs in the abdomens, in addition to tymbal muscles.
In the experiments, we investigated the suitable voltage for control cicadas and the limitation of the generated sound expressions of the cicada speaker.
Our contributions are as follows.

\begin{itemize}
\item We demonstrate manipulation of the musical scales of the chirps of cicadas by using electrical muscle stimulation.
\item For producing sounds from a cicadas, we conducted experiments on the correlation between input voltages and the types of generated sounds, and discussed the limitation of generated sounds with our methods.
\end{itemize}

\begin{figure*}[t]
\centering
\subfigure[]{\includegraphics[width=0.9\textwidth]{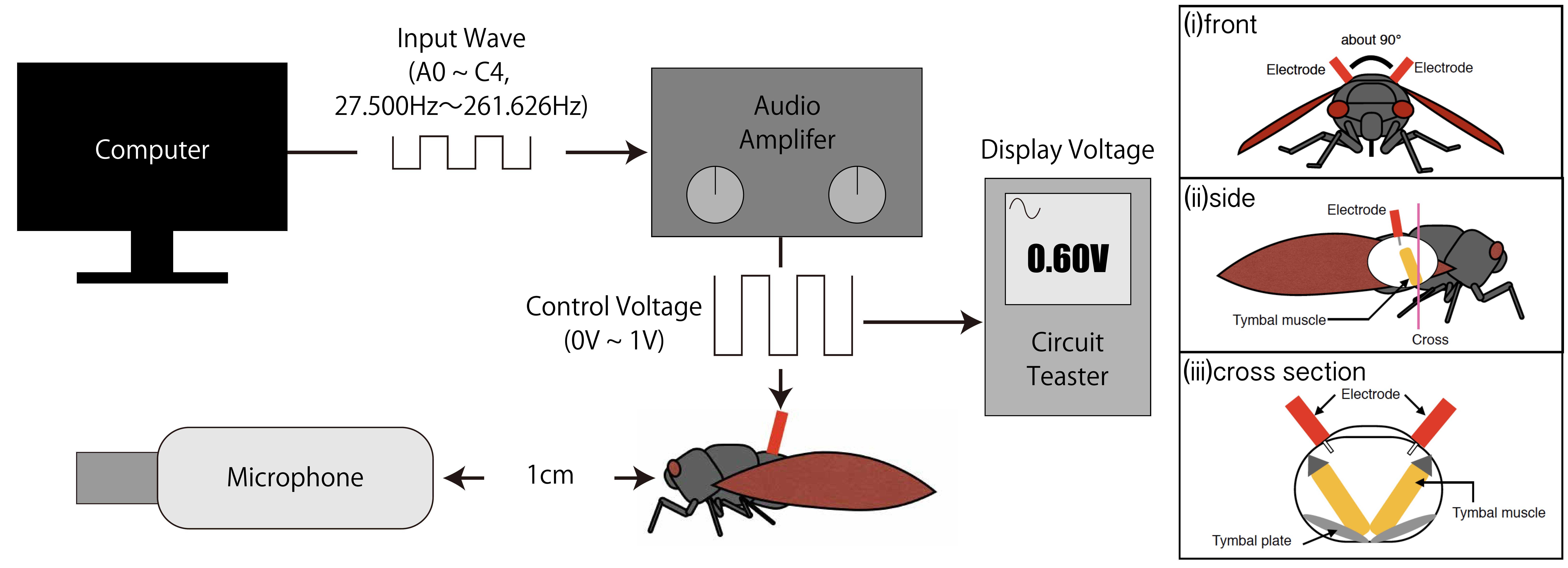}}
\subfigure[]{\includegraphics[width=0.9\textwidth]{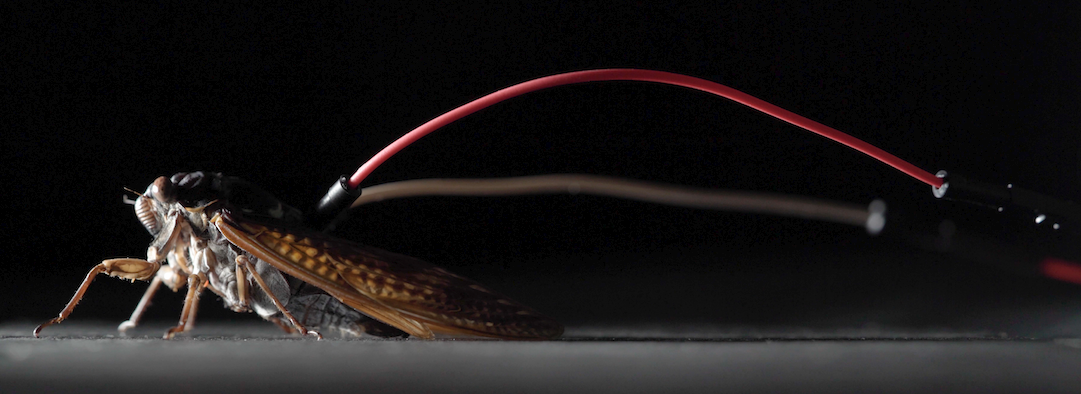}}
\caption{
In the left of (a), our user interface of controlling the pitch of a cicada for producing voices by using electrical stimulation on the tymbal muscles of the cicada.
In the right of (a) we show the inserted electrodes into a cicada for controlling as follows.
(i) The view from the front of the cicada inserted the electrodes. 
(ii) The view from the side of the cicada inserted the electrodes and abdominal part near the tymbal muscles. 
(iii) A cross section of the cicada that inserted the electrodes, which is from the pink line.
In (b), we show the image of electrodes attached on a cicada.}\label{fig:system}
\end{figure*}

\section{Insect-Computer Interface: Controlling cicada sound by electrical muscle stimulation, at a specific pitch interval}

In Figure \ref{fig:system}, we use the male cicada of Graptopsaltria nigrofuscata (a large brown cicada), which has the following advantages.
\begin{itemize}
\item The cicada has fewer muscles or organs near the tymbal muscles (used to produce sound) in the body, compared to the tymbal muscles of other insects, which makes electrical stimulation of the tymbal muscles of a cicada easier.
\item The \textit{Graptopsaltria nigrofuscata cicada} is a relatively large species and can be easier to perform surgical procedures such as electrode insertions. In addition, only the male can chirp.
\end{itemize}

We demonstrate the surgery to make a cicada to generate the chirping waveform by using electrical stimulation as follows. Chirping is produced by contracting the tymbal muscles to vibrate the tymbal plates and amplify the sound in the abdominal cavity \cite{Bennet-Clark1681, Popov1981, Young1001}. 
Therefore, the chirp can be controlled by electrical stimulation of the tymbal muscles. 
The process is as follows. 
The electrodes are inserted into the abdomen of the cicada through the back \cite{DBLP:journals/computer/BozkurtLS16}, as shown in Figure \ref{fig:system}. 
In addition, the other side of the electrode is connected to a circuit for signal input. 
The waveforms generated to input to a cicadas are square waves, with frequencies from $A0$ to $C4$ ($27.500 Hz$ to $261.626 Hz$).

% \begin{figure*}[t]
% \centering
% \includegraphics[width=0.5\textwidth]{fig/system.png}
% \caption{The user interface of how we control Cicada for }\label{fig:system}
% \end{figure*}

\section{Experimental Results}

We experiment on the mechanism of control to produce sounds from a cicada by our interface.
Specifically, we analyze which range of input voltage (for example, the range of values from $0 V$ to $1 V$) is required to produce the sound of a cicada, given a specific input interval of pitch (for example, $C\#3$).

\subsection{Experimental Setting}

As shown in our interface in Figure \ref{fig:system}, we show how a cicada is electrically stimulated and how we measured the sound of the produced cicada and related input signals.
The interface sends an electric signal through the amplifier circuit, and the amplifier circuit enhances the signal and sends it to the semi-sounding muscle of a cicada.
We set a microphone located in front of the cicada with one centimeter to record the cicada's sound, when the cicada was electrically stimulated. 
Note that we used the ECM-CG60\footnote{\url{https://www.sony.jp/handycam/products/ECM-CG60/}} (SONY) as a microphone, whose frequency response range is $40 Hz$ to $20 kHz$. 
In addition, we fixed the wings of the cicada to keep the same distance between the microphone and the cicada.

\textbf{Measurement of a process}.
We installed the circuit tester in parallel on the cicadas for measuring voltage which was flowing through the cicada's body when the cicadas were making sounds. 
Specifically, we generated a square wave of frequency from $A0$ to $C4$ ($27.500 Hz$ to $261.626 Hz$).
In a process, given a specific input waveform of frequency (e.g. $A0$), we gradually increased the voltage from $0 V$ to $1 V$, and recorded down if the cicada started and continued to produce sound.
We terminated the process until the cicada stopped producing sound.
Note that the voltage value was displayed on the tester.
In the experiments, there were seven cicadas captured in Tsukuba, Japan, and we repeated the same process on each cicada, for the generated waveforms from A0 to C4.

%The electricity of a certain frequency was continuously passed through the semi, and the voltage value was gradually increased from 0V. 
%Then we stopped the electrical stimulation when the voltage was too high and the cicadas stopped ringing. 

% 図はセミに電気刺激をし鳴き声と入力信号を計測する環境である。 コンピュータからアンプ回路に電気信号を送り、 増幅したものをセミの発音筋に流す。アンプ回路にはセミと並列にテスターが取り付けられており、 入力信号のデバックを行う事と同時に電圧値を計測する事ができる。以下で、 各部分の詳細な説明を行う。
% 蝉が電気刺激を受けたときに蝉の鳴き声を録音するためのマイクとしてECM-CG60(SONY)を使用した。このマイクの周波数特性の範囲は40Hz〜20kHzである。マイクは1cmの距離を開いて蝉の前に位置した(図2)。

% 両翼が固定されていない状態で蝉が羽ばたくと、 マイクとの間隔と位置がずれてしまう。 そのため蝉の羽は固定されていた。 セミが鳴いているときに、どれだけの電圧が蝉に流れているかを測定するために、 (回路)テスタが蝉と並列に回路に組み込まれた。 電圧を記録して、 セミにかかる電圧を確認した。 

% また、 テスターによって流れている電気の周波数を確認することでデバッグした。
% 前述のように、 周波数がA0〜C4(27.500Hz〜261.626Hz)の矩形波を生成した。 コンピュータからアンプ回路を通じて、 セミの発音筋に電気刺激を与えていく。 一定の周波数の電気をセミに流し続け、 電圧値を0Vから徐々に上げていった。 それから私達は電圧値が上がり過ぎ、 セミが鳴かなくなった時、 私達は電気刺激を止めた。 テスターに電圧値を表示し、 セミの鳴き声とセミにかかっている電圧の関係を記録した。 A0からC4まで同じプロセスを繰り返した。

\begin{figure}[t]
\centering
\includegraphics[width=0.9\textwidth]{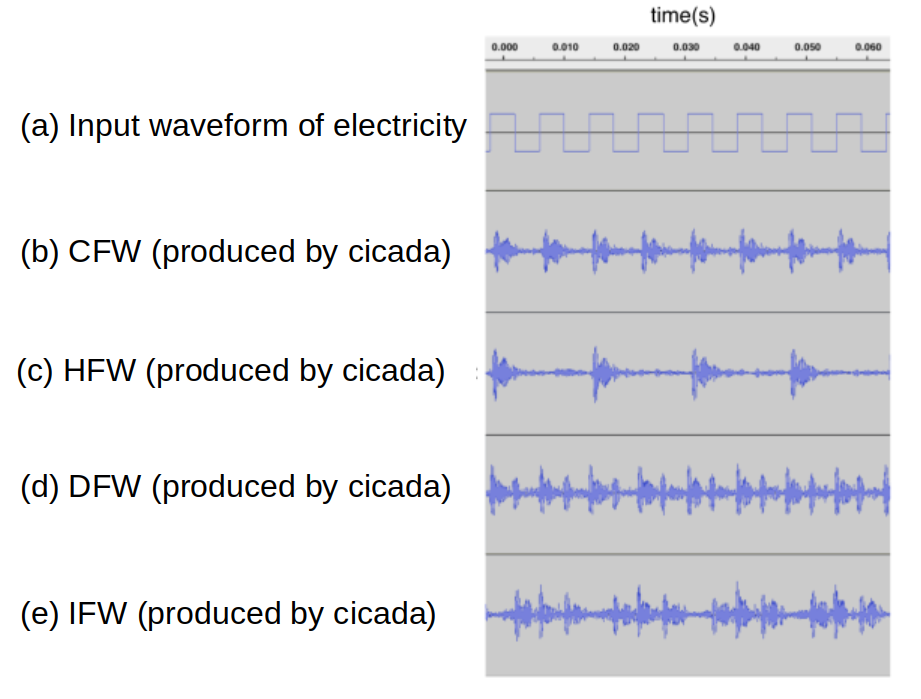}
\caption{With a specified frequency of electrical stimulation (i.e., interval of pitch),
we showed that (a) the input electrical waveform, and the following four are produced by cicada's sound, 
which are (b) the waveform of $CFW$ (Correct Frequency Wave), 
(c) the waveform of HFW (Half Frequency Wave),
(d) the waveform of DFW (Double Frequency Wave),
and the waveform of IFW (Irregular Frequency Wave).}\label{fig:wave}
\end{figure}

\begin{figure}[t]
\centering
\includegraphics[width=0.9\textwidth]{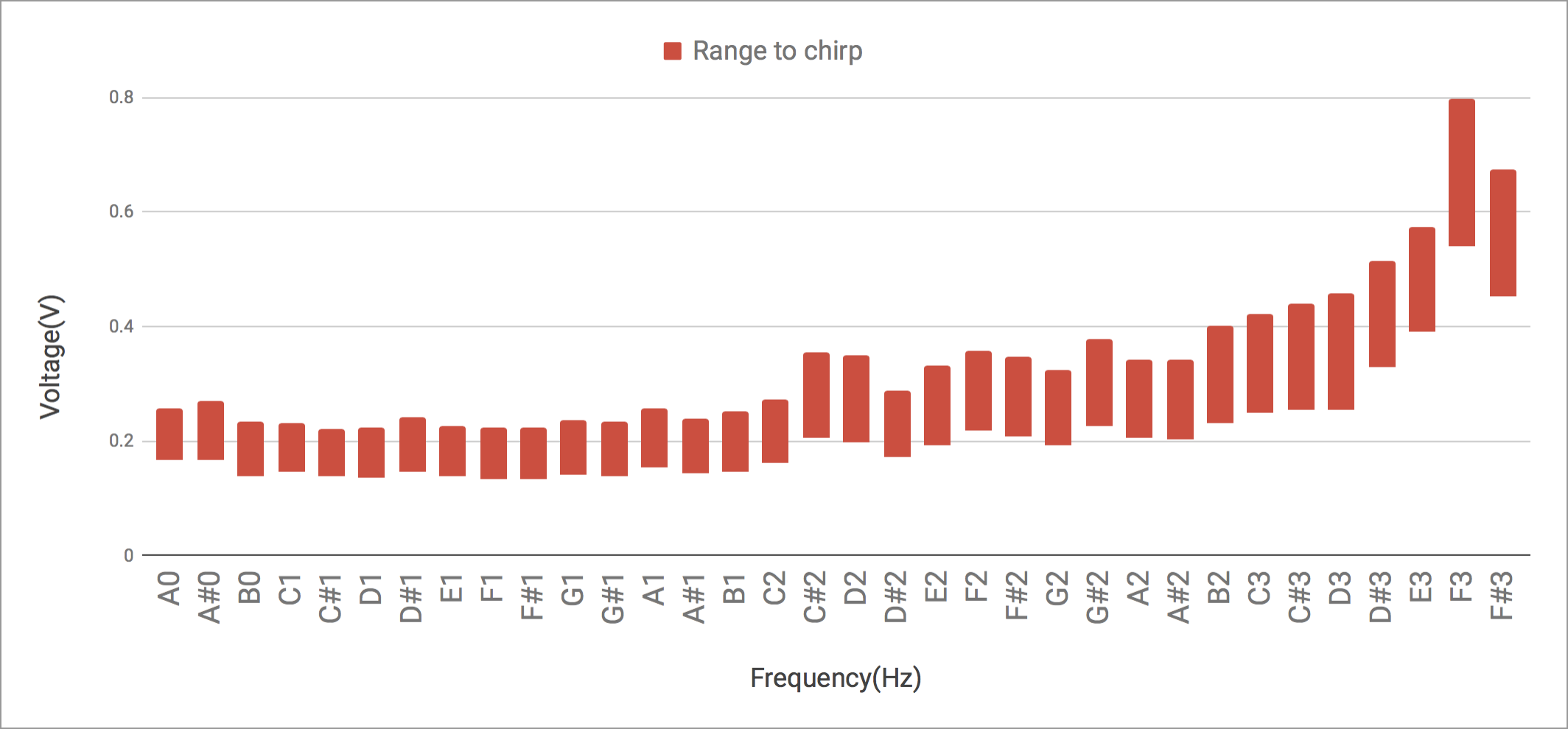}
\caption{In experiments, we analyzed the range of input voltages were required for producing cicada's sound for each interval of pitch. 
We found that the type of the cicada's sound, which had the CFW waveform (i.e., those having similar interval with the input electrical waveform), could be produced in most intervals of pitches (i.e., from $A0$ to $F\#3$).}\label{fig:voltage_frequency_cfw}
\end{figure}

\begin{table}[t]
\centering
\scriptsize

\caption{The frequency of interval of pitch and the maximal pitch cicadas could produce.}
\begin{tabular}{|c|c|c|c|c|c|c|c|c|c|c|c|}
\hline
                                                                                         & G\#2    & A2  & A\#2    & B2      & C3      & C\#3    & D3      & D\#3    & E3      & F3      & F\#3    \\ \hline
Frequency                                                                                & 103.826 & 110 & 116.541 & 123.471 & 130.813 & 138.591 & 146.832 & 155.563 & 164.814 & 174.614 & 184.997 \\ \hline
\begin{tabular}[c]{@{}c@{}}Number of maximal \\ pitch cicadas could produce\end{tabular} & 1       & 1   & 0       & 0       & 1       & 2       & 0       & 0       & 1       & 0       & 1       \\ \hline
\end{tabular}
\end{table}

\subsection{Analyzing among Input Waveforms (Frequencies), Input Voltages, and the Produced Waveforms (Frequencies) of Cicada's Sound}

During the experiment of each cicada, we classed the waveform generated by cicada chirps into four patterns and thus investigated which cicada's waveform pattern was appropriate in most of intervals of pitches.
We practically played Pachelbel's canon\footnote{https://en.wikipedia.org/wiki/Pachelbel's Canon} and examined whether the chirps of the cicada were produced. 
In Figure \ref{fig:wave}, about the waveforms of cicada chirps, the HFW could be produced when entering the voltage lower than the voltage of producing the CFW. Similarly, DFW could be produced by inputting a voltage higher than the voltage of producing CFW.

Note that in an experiment, on a specific input electrical frequency, when gradually increasing the voltage to a cicada, we could observe a possible appearing order of HFW, CFW, and DFW sequentially. 
However, under a quite low frequency of electricity (that is, low pitch interval), a cicada was not stimulated to generate the HFW of the sound.
Furthermore, under a quite high frequency of electricity (that is, a high interval of pitch), a cicada was not stimulated to generate the DFW of the sound.

Specifically, in Figure \ref{fig:voltage_frequency_cfw}, in the experiments of seven cicadas, we showed the range of the input voltages that triggered the CFW waveform of the cicada sound, and we observed that the voltages of CFW could cover more range of intervals of pitches, than those voltages of HFW or DFW.
The range covered by the CFW waveform of the cicada sound was from A0 to F\#3.
Note that the mean maximum frequency of the chirps of the cicada was the pitch of C \#3 (13.891 Hz).

\section{Conclusion}

% Conclusion goes here.

We proposed a method to manipulate the musical scales of cicadas' chirps by using electrical muscle stimulation (EMS).
We evaluated cicadas as speakers with the suitable range of voltage for control cicadas' chirps and the limitation of the generated chirp expressions.
Our experimental results shed light on the production of insect-based communication tools that are energy-efficient, durable, and agile, compared to common robots. 
In the future, we expect these insect-based communication tools to be used as audio supports in emergency situations such as disasters.

% We investigated a novel method to manipulate the chirps of cicadas with EMS to send music messages that can be perceived by people. 
%Our main contributions are below.
% The possibility of interactions between people by using chirps of insects (i.e., cicadas). 

\bibliographystyle{ACM-Reference-Format}
\bibliography{main}

\end{document}